\documentclass[prl,twocolumn,showpacs,preprintnumbers,amsmath,amssymb]{revtex4}

\usepackage{graphicx}
\usepackage{dcolumn}
\usepackage{bm}

\begin{document}


\title{Unconventional Superconductivity and Electron Correlations in Cobalt Oxyhydrate Na$_{0.35}$CoO$_{2}$$\cdot y$H$_{2}$O}

\author{Tatsuya~Fujimoto$^1$}
\author{Guo-qing~Zheng $^1$}%
\altaffiliation[author to whom correspondence should be addressed. ]{e-mail: zheng@mp.es.osaka-u.ac.jp.}
\author{Y.~Kitaoka $^1$}%
\author{R.L.~Meng $^2$}%
\author{J.~Cmaidalka $^2$}%
\author{C.W.~Chu $^{2,3,4}$}%

\address{$^1$ Division of Frontier Materials Science, Graduate School of Engineering Science, Osaka University,  Osaka 560-8531, Japan}
\address{$^2$ Department of Physics and TCSAM, University of Houston, TX 77204-5932.}
\address{$^3$ Lawrence Berkeley National Laboratory, 1 Cyclotron Road, Berkeley, CA 94720} 
\address{$^4$ Hong Kong University of Science and Technology, Hong Kong}


\begin{abstract}
We report a precise  $^{59}$Co nuclear quadrupolar resonance (NQR) measurement on the recently discovered cobalt oxyhydrate Na$_{0.35}$CoO$_{2}$$\cdot y$H$_{2}$O superconductor from $T$=40 K down to 0.2 K. We find that in the normal state the spin-lattice relaxation rate $1/T_1$ follows a Curie-Weiss type temperature ($T$) variation, $1/T_1T=C/(T-\theta)$, with $\theta$=-42 K, suggesting two-dimensional antiferromagnetic spin correlations. Below $T_c$=3.9 K, $1/T_1$ decreases with no  coherence peak  and follows a $T^n$ dependence with $n\simeq$2.2 down to $\sim$2.0 K but crosses over to a $1/T_1\propto T$ variation below $T$=1.4 K, which suggests non s-wave superconductivity. The data in the superconducting state are most consistent with the existence of line nodes in the gap function.   
\end{abstract}

\maketitle

\sloppy

The recent discovery of superconductivity in layered cobalt oxyhydrate Na$_{0.35}$CoO$_{2}$$\cdot y$H$_{2}$O \cite{Takada} has generated new excitement in the condensed matter physics community,  since it suggests a possible new route to find high-$T_c$ superconductivity and also demonstrates  the richness of physics in layered transitional metal oxides. The compound consists of two-dimensional CoO$_2$ layers separated by insulating blocks of Na$^{+1}$ and H$_2$O molecules, resembling the layered structure of copper-oxide high-$T_c$ superconductors.
 Because of the octahedral crystal environment, Co$^{+4}$ is in the low spin state ($S$=1/2), and the compound Na$_{0.35}$CoO$_{2}$$\cdot y$H$_{2}$O is considered as a system in which 35\% electrons are doped to a spin 1/2 triangular lattice. Insight from investigations of this compound is expected to shed light on the mechanism of cuprate superconductors. Several experiments \cite{Sakurai,Lorenz,Schaak,Cao,Jin} have been conducted to investigate its physical properties and many theoretical proposals on the symmetry of the superconductivity \cite{Baskaran,Kumar,WangQH,Ogata,Hu,Zhang} have been put forward.

In this Letter, we report a precise measurement using $^{59}$Co nuclear quadrupolar resonance (NQR) on the normal and superconducting states of Na$_{0.35}$CoO$_{2}$$\cdot y$H$_{2}$O. In the normal state, it was found that the spin lattice relaxation rate $1/T_1$ divided by temperature ($T$) follows a Curie-Weiss law of $1/T_1T \propto 1/(T+42)$, suggesting the two dimensional antiferromagnetic spin correlations. In the superconducting state, $1/T_1$ decreases below $T_c$=3.9 K, with no coherence peak and follows a power-law of  $T^n$ with $n\simeq$2.2 down to $\sim$2.0 K but crosses over to a $1/T_1\propto T$ variation below $T$=1.4 K, which suggests unconventional superconductivity. These results show that the layered Co compound bears a close resemblance to the Copper-oxide high-$T_c$ superconductors without a full gap.

The Na$_x$CoO$_2$$\cdot$yH$_2$O powder was synthesized as described in Ref. \cite{Takada} and \cite{Meng}. The x-ray spectrum shows the reflections of the hexagonal space group P63/mmc with lattice parameters a=2.820 $\AA$ and c=19.593 $\AA$ \cite{Lorenz}. $T_c$ was found to be 3.9 K from the ac susceptibility measured by using the in-situ NQR coil. NQR measurements were carried out using a phase-coherent spectrometer. Measurements below 1.4 K were performed by using a $^3$He/$^4$He dilution refrigerator and a small amplitude of radio-frequency pulses. The  $1/T_1$ of $^{59}$Co was measured by the saturation-recovery method. Three NQR transition lines arising from spin $I$=7/2 of $^{59}$Co    corresponding to $\nu_Q$=4.12 MHz and the asymmetry parameter $\eta$=0.223 were found. Here $\nu_{Q}$ and $\eta$ are defined as $\nu_{Q}\equiv\nu_{z}=\frac{3}{2I(2I-1)h}e^2Q\frac{\partial ^2V}{\partial z^2}$, and $\eta=|\nu_{x}-\nu_{y}|/\nu_{z}$,  with $Q$ being the nuclear quadrupolar moment, $I$=7/2 being the nuclear spin and $\frac{\partial ^2V}{\partial \alpha^2} (\alpha=x, y, z)$ being the electric field gradient at the position of the nucleus \cite{Abragam}. The inset of Fig. 1 shows the $\pm$ 3/2$\leftrightarrow$$\pm$5/2 transition line at whose peak $T_1$ was measured. The spectrum has a full-width-at-half-maximum of 0.3 MHz. The main panel of the figure shows the typical nuclear magnetization $M(t)$ that is excellently fitted to the theoretical curve $\frac{M_0-M(t)}{M_0}=0.095exp(-3t/T_1)+0.095exp(-9.5t/T_1)+0.819exp(-19t/T_1)$ \cite{Mac}, with a unique $T_1$ component. Both the fairly narrow NQR spectrum and the single-component nuclear magnetization curve  indicate good quality of the sample.

\begin{figure}
\begin{center}
\includegraphics[scale=0.5]{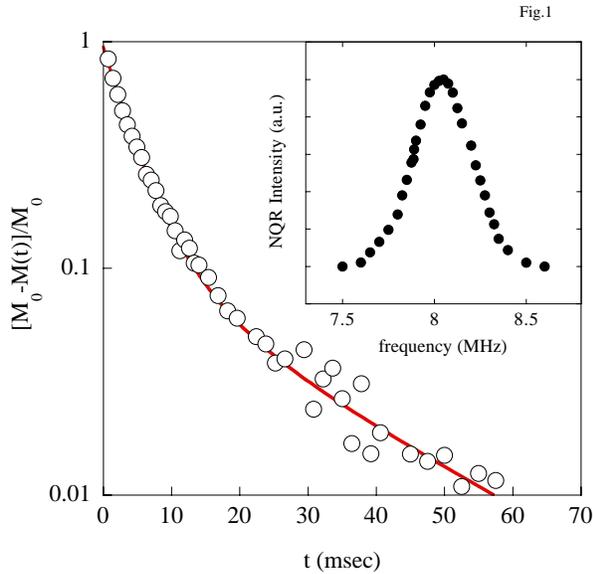}
\caption{$^{59}$Co nuclear magnetization decay curve at $T$=1.7 K fitted uniquely by the expected theoretical curve (see text). The inset shows NQR line shape ($\pm 3/2 \leftrightarrow \pm$5/2 transition) at whose peak $T_1$ was measured.}
\label{fig:1}
\end{center}
\end{figure}

Figure 2 shows the temperature dependence of $1/T_1$. Below $T_c$, $1/T_1$ decreases with no coherence (Hebel-Slichter) peak, and follows a power-law dependence of  $T^n$.  If one fits the data for 3.8 K$\geq T\geq$2.0 K, one gets an $n\simeq$2.2.  This is strong evidence for non-s wave superconductivity. For s-wave, isotropic gap, $1/T_1$ would show a coherence peak just below $T_c$ and follows an exponential temperature dependence at lower temperature. The most immediate explanation of our data is the possible existence of line nodes in the gap function  as in $d$- or $p$-wave superconductors, where  an energy ($E$)-linear density of states (DOS) at low $E$  results in a $T^n$ ($n$=3) dependence of $1/T_1$. In the presence of disorder/impurity scattering, the exponent $n$ could appear smaller than 3 as shown shortly. For $d$- or $p$-wave superconductivity, the "coherence factor" \cite{Schrieffer} due to isotropic pairing is absent, therefore the enhancement of $1/T_1$ just below $T_c$  is greatly reduced. If there is no strong divergence of the DOS at $E=\Delta_0$ ($\Delta_0$ is the maximum gap amplitude), there will be no peak seen just below $T_c$. The slight retarded (by 0.1 K in temperature) drop of $1/T_1$ below $T_c$ in the present case could arise from two possibilities. First, the gap amplitude $\Delta_0$ is small, as shown by a simple simulation described below. Second, bulk superconductivity sets in at a temperature 0.1 K below the onset of diamagnetism seen in the susceptibility, which is the case that has been encountered previously in some superconductors. For example, in the heavy fermion superconductor CeIrIn$_5$, although the susceptibility shows a superconducting onset at $\sim$0.6 K \cite{Petrovic}, $1/T_1$ only starts to drop  at 0.4 K \cite{Zheng}. This also happens in the isostructure heavy fermion compound CeRhIn$_5$ in the vicinity of magnetic order where $1/T_1$ shows a rapid decrease at a temperature far below the susceptibility onset temperature \cite{Kawasaki}. 

\begin{figure}
\begin{center}
\includegraphics[scale=0.5]{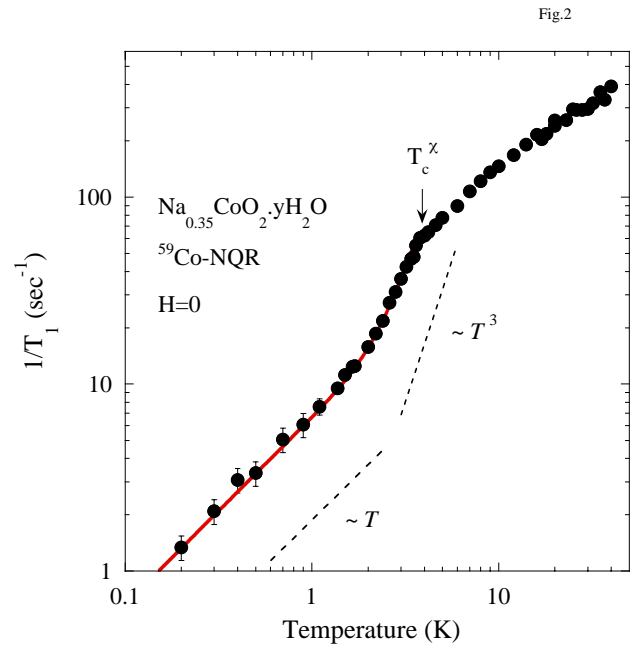}
\caption{$1/T_1$ as a function of temperature. The arrow indicates the superconducting onset temperature in the susceptibility.  The curve below $T_c$ is a calculation for the line-nodes gap function in the presence of impurity scattering (see text for detail).}
\label{fig:2}
\end{center}
\end{figure}

At lower temperatures, the decrease of $1/T_1$ becomes more gradual, and below $T$=1.4 K $1/T_1$ becomes to be proportional to $T$ down to $T$=0.2 K. In the case of nodal superconductivity, this result can then be interpreted as due to disorder or impurity that acts as pair breaker \cite{Miyake,Hotta}. $T_1$ in the superconducting state can be expressed as 
$\frac{T_1(T=T_c)}{T_{1}} = \frac{2}{k_BT_c}\int (\frac{N_{s}(E)}{N_0})^{2}f(E)(1-f(E))dE$, 
where $\frac{N_{s}(E)}{N_{0}}=\frac{E}{\sqrt{E^{2}-\Delta^{2}}}$  with $N_{0}$ being the DOS in the normal state and
$f(E)$ being the Fermi function.    
The solid curve below $T_c$ shown in Fig. 2 is a calculation assuming line nodes in the gap function, $\Delta=\Delta_0 cos2\theta$, with $\Delta_0$=2.6$k_BT_c$ and a residual DOS ($N_{res}$) due to impurity (disorder) scattering \cite{Hotta}, $N_{res}$=0.65$N_{0}$. The curve fits the data quite well. The deviation from the $T^3$ law at quite a high temperature is hence an effect of the large $N_{res}$ that produces a large value of $1/T_1\propto T$. The gap function of $\Delta=\Delta_0 cos\theta$, with $\Delta_0$=3.0$k_BT_c$ and  $N_{res}$=0.65$N_{0}$ can also fit the data. It should be emphasized that  s-wave or extended s-wave superconductivity in the presence of impurity scattering will lead to a fully opened gap \cite{Norman}, which is inconsistent with our results. Finally, we note that it has been shown that the anisotropic triangular lattice Hubbard model at or near half filling yields superconductivity with a line-nodes gap ($d$-wave gap) \cite{Dagotto,Kontani}.

 The clean chiral d-wave or p-wave superconductivity with a full gap  \cite{Baskaran,Kumar,WangQH,Ogata,Hu} does not find immediate agreement with our data. On the other hand,  the presence of disorder or impurity may fill up the gap, rendering the low-$E$ DOS. Very recently,  Wang and Wang calculated the effect of a single impurity/interface on $d_{x^2-y^2}+id_{xy}$ or $p_x+ip_y$ superconductivity \cite{ZDWang}, and they found that the impurity-induced sub-gap states tend to cover the whole gap regime of the chiral superconductivity. However, it would appear to be a coincidence that an $E$-linear DOS shows up in such a case. Measurements in samples with different degree of defect or Na concentration may help settle this issue. On the theoretical side, a detailed calculation of the spatially-averaged DOS for finite concentration of impurity is desirable for comparison with experiments.

Next, we turn to the normal state. In Fig. 3 is shown the temperature variation of $1/T_1T$. There, it is seen that the $1/T_1T$ increases with decreasing $T$. This is a feature not seen in conventional metals where  $1/T_1T$ would be a constant, but resembles the Cu relaxation in cuprate high-$T_c$ superconductors with layered square lattices. Above $T$=10 K, $1/T_1T$ follows a Curie-Weiss law, namely $1/T_1T=\frac{C}{T-\theta}$, with $\theta$=-42 K and $C$=750 Sec$^{-1}$.  We interpret this feature as arising from two dimensional antiferromagnetic fluctuations. For antiferromagnetically correlated itinerant electron systems, the staggered spin susceptibility is shown to follow a Curie-Weiss type temperature variation \cite{Millis,Moriya}. Since $1/T_1T$ probes the dynamical susceptibility averaged over the momentum space, it is a good probe of electron correlation, as has been proven in cuprate superconductors. In the present case, although ferromagnetic correlation is present for large Na concentration ($x$=0.7) \cite{Motohashi}, antiferromagnetic coupling was found in low concentration of Na \cite{Wang} and also in Na$_{0.35}$CoO$_{2}$$\cdot y$H$_{2}$O \cite{Takada}.  
 
 \begin{figure}
\begin{center}
\includegraphics[scale=0.5]{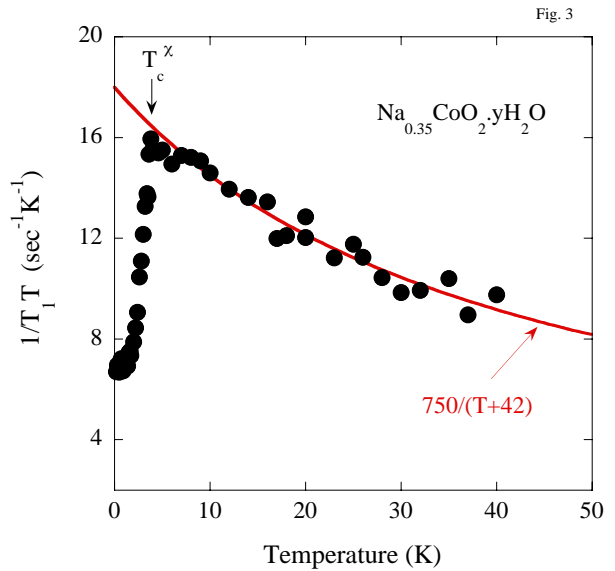}
\caption{$1/T_1T$ develops with decreasing temperature. The curve above $T$=10 K is a fitting to $C/(T+\theta)$.}
\label{fig:3}
\end{center}
\end{figure}

The unconventional superconductivity developed with the background of antiferromagnetic spin correlations in Na$_{x}$CoO$_{2}$$\cdot y$H$_{2}$O bears close similarities to the case of high-$T_c$ cuprates and suggests the importance of electron correlations in the occurrence of the superconductivity in the layered electron systems. In this regard, the lower $T_c$ of the cobaltate may be ascribed to its antiferromagnetic exchange energy that is smaller by one order in magnitude than in cuprates \cite{Wang}.

In conclusion, we have presented a  $^{59}$Co NQR study on the recently discovered cobalt oxyhydrate Na$_{0.35}$CoO$_{2}$$\cdot y$H$_{2}$O superconductor from $T$=40 K down to 0.2 K. In the normal state the spin-lattice relaxation rate $1/T_1$ follows a Curie-Weiss type temperature ($T$) variation, $1/T_1T=C/(T-\theta)$, with $\theta$=-42 K, suggesting two-dimensional antiferromagnetic spin correlations. Below $T_c$=3.9 K, $1/T_1$ decreases with no apparent coherence peak, and follows  $T^n$ ($n\sim$2.2) dependence down to 2.0 K. At lower temperatures, the decrease of $1/T_1$ becomes more gradual and a $1/T_1\propto T$ variation was observed below $T$=1.4 K. These data suggest non s-wave superconductivity, and are most consistent with the line-nodes gap model in the presence of scattering by disorder/impurity. 
None of current theories proposed for the gap symmetry of the CoO$_2$-based superconductor finds immediate agreement with our results, and we hope our experiment will stimulate further theoretical works.

After completing this work, we became aware of two similar studies by the authors of Ref. \cite{Kobayashi} and Ref. \cite{Yoshimura}, but their nuclear magnetization curves can not be fitted by a unique $T_1$  and the inferred $1/T_1$ results that show a coherence peak, are quite different from ours.

We thank  K. Miyake and H. Kohno for  helpful discussions and A. V. Balatsky and Z. D. Wang for  correspondence on the impurity effect in chiral superconductors. This work was supported in part by research grants from MEXT (Nos. 14540338 and 15GS0213), NSF (No. DMR-9804325), the T. L. L. Temple Foundation, the John abd Rebecca Moores Endowment, and DoE (No. DE-AC03-76-SF00098).


\end{document}